\documentclass[cite,doublecol,figures]{epl2}
\usepackage{epsfig,color}
\usepackage{amsfonts}
\usepackage{amssymb}
\usepackage{amsmath}
\usepackage{verbatim}
\definecolor{brblue}{rgb}{0,1,1}
\definecolor{orange}{rgb}{1,0.5,0}
\usepackage[english]{babel}

\newcommand{\mb}{\mathbf}
\newcommand{\Litwo}{Li$_2$CuO$_2$}

\newcommand{\K}{\,\mathrm{K}}

\newcommand{\meV}{\,\mathrm{meV}}
\newcommand{\Angstrom}{\,\mathrm{\AA}}

\newcommand{\TN}{T_\mathrm{N}}
\newcommand{\Hamilton}{\textsc{Hamilton}}
\newcommand{\Heisenberg}{\textsc{Heisenberg}}
\newcommand{\Hubbard}{\textsc{Hubbard}}
\newcommand{\Neel}{\textsc{N\'eel}}

\newcommand{\IFW}{Leibniz-Institut f{\"u}r Festk\"orper- und Werkstoffforschung (IFW) Dresden, Germany\\}
\newcommand{\IFP}{Institut f\"ur Festk\"orperphysik, Technische Universit\"at Dresden, 01062 Dresden, Germany\\}
\newcommand{\ILL}{Institut Laue Langevin, F-38042 Grenoble Cedex 9, France\\}
\newcommand{\JCNS}{J\"ulich Centre for Neutron Science JCNS, J\"ulich, Germany\\}
\newcommand{\IMSKiev}{Institute for Problems of Materials Science Krzhizhanovskogo 3, 03180 Kiev, Ukraine\\}
\newcommand{\MPICPfS}{Max-Planck-Institut f\"ur Chemische Physik fester Stoffe, Dresden, Germany\\}
\newcommand{\Prague}{Institute of  Physics, ASCR, Prague, Czech Republic\\}

\def\tf {$T_{\rm 2}$}

\newcommand{\figref}[1]{Fig.\,\protect\ref{#1}}
\newcommand{\eqnref}[1]{Eqn.\,(\protect\ref{#1})}
\newcommand{\tabref}[1]{Table\,\protect\ref{#1}}

\title{Highly Dispersive Spin Excitations in the Chain Cuprate \Litwo}

\shorttitle{Spin excitations in \Litwo }

\author{W.E.A.\ Lorenz\inst{1}\thanks{E-mail: \email{w.lorenz@ifw-dresden.de}} \and R.O.\ Kuzian\inst{1,2}
\and S.-L.\ Drechsler\inst{1} \and W.-D.\ Stein\inst{3} \and N.\ Wizent\inst{1} \and G.\ Behr\inst{1}
\and J. M\'alek\inst{1,7} \and U.\ Nitzsche \inst{1} \and  H.\ Rosner\inst{6} \and A.\ Hiess\inst{4} \and W.\ Schmidt\inst{5}
\and R.\ Klingeler\inst{1} \and M.\ Loewenhaupt\inst{3} \and B.\ B\"uchner\inst{1} \shortauthor{\sc
W.E.A.\ Lorenz \etal }}

\institute{ \inst{1}\IFW \inst{2}\IMSKiev \inst{3}\IFP \inst{4}\ILL \inst{5}\JCNS \inst{6}\MPICPfS
\inst{7}\Prague

}


\abstract{We present an inelastic neutron scattering investigation of
\Litwo \ detecting the long sought quasi-1D
magnetic excitations with a large dispersion along the CuO$_2$-chains 
studied up to  
$25 \meV$.   
The total dispersion is governed by a surprisingly 
large ferromagnetic (FM) nearest-neighbor
exchange integral $J_1=-228$~K.
An anomalous
quartic dispersion near the zone center and a pronounced minimum
near (0,0.11,0.5) r.l.u.\ 
(corresponding to a spiral excitation with a pitch angle about 41$^{\circ}$) point
to the vicinity of a 3D FM-spiral critical
point. The leading exchange couplings are obtained
applying standard linear spin-wave theory.
The 2$^{nd}$ neighbor inter-chain interaction 
suppresses a spiral state and drives the FM
in-chain ordering below the \Neel{} temperature. The obtained
exchange parameters are in agreement with the results for a realistic five-band
extended \Hubbard{} Cu $3d$ O $2p$ model and LSDA+$U$ predictions. }

\begin{document}

\maketitle

\section{{\bf 1. INTRODUCTION\/}}
Li$_2$CuO$_2$ is the first \cite{Hoppe1970} and the most frequently studied compound of the
growing class of edge-shared spin-chain cuprates
\cite{Matsuda1996,Matsuda2001,MasudaPRB2005,Enderle2005,Drechsler2007PRL,Drechsler2007JP}. Owing to its structural
simplicity with ideally planar CuO$_2$ chains (see Fig.\ 1) it has been considered as a model
quasi-one-dimensional (1D) frustrated quantum spin system. In almost all edge-shared cuprate chain
compounds the spins are expected to be coupled along the chains via nearest neighbor (NN)
ferromagnetic (FM) and next-nearest neighbor (NNN) antiferromagnetic (AFM) exchange interactions,
$J_1$ and $J_2$, respectively. Due to the induced frustration FM and spiral in-chain correlations
are competing. While in the 1D model the ground state is governed by the ratio $\alpha=-J_2/J_1$,
the actual 3D magnetic order sensitively depends on the strength of the inter-chain couplings and
anisotropy. In \Litwo{},  below $\TN \approx 9 \K$ \cite{Sapina1990,Chung2003}
a long-range collinear commensurate
AFM inter-chain with  FM in-chain (CC-AFM-FM) 
magnetic ordering evolves.
However, a
proper 
understanding, necessary for
\begin{figure}[!b]
\begin{center}
\begin{minipage}{0.9\textwidth}
\hspace{.03\textwidth}
\includegraphics[width=0.13\textwidth]{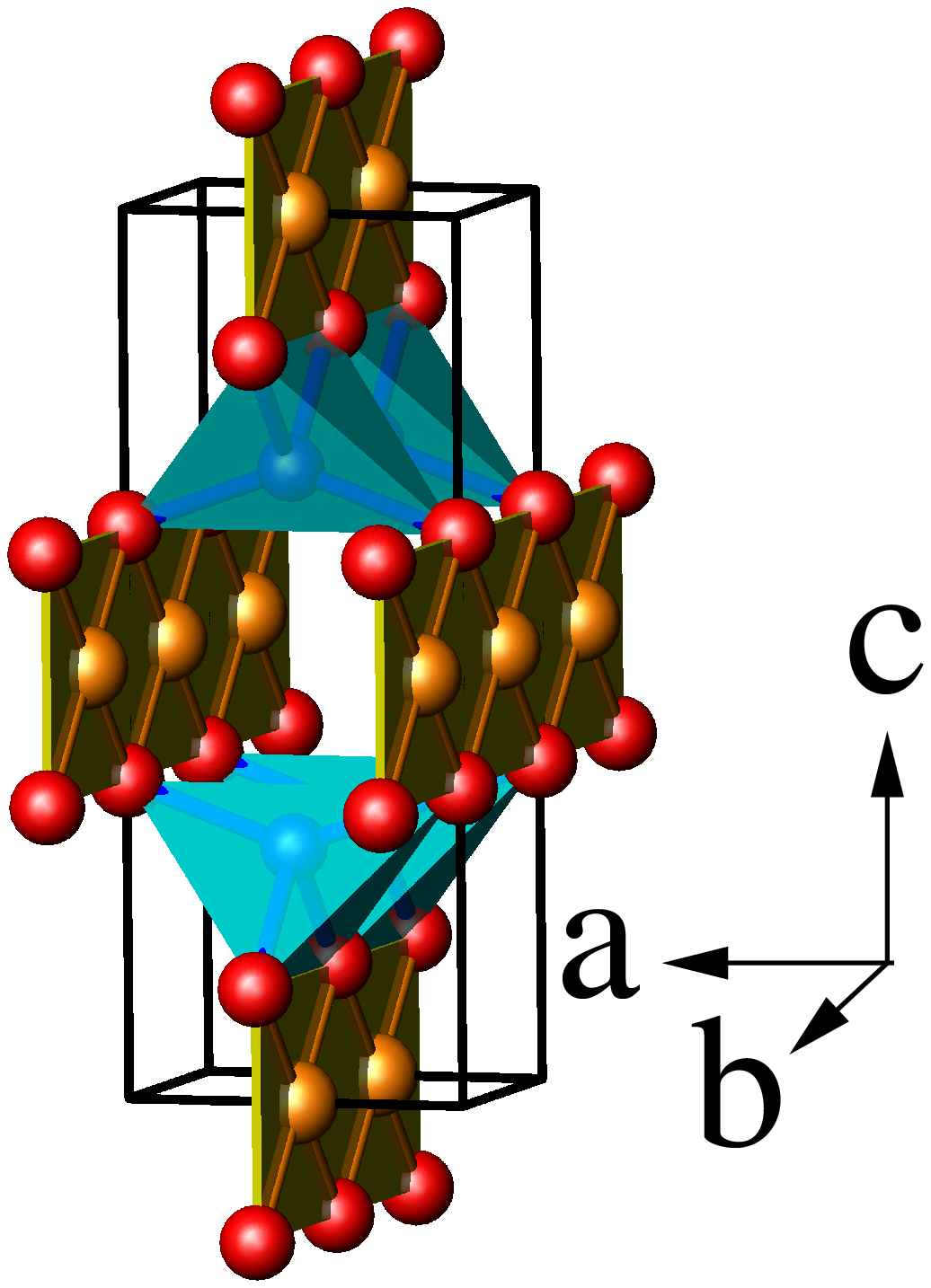}
\hspace{.005\textwidth}
\includegraphics[width=0.35\textwidth,angle=0]{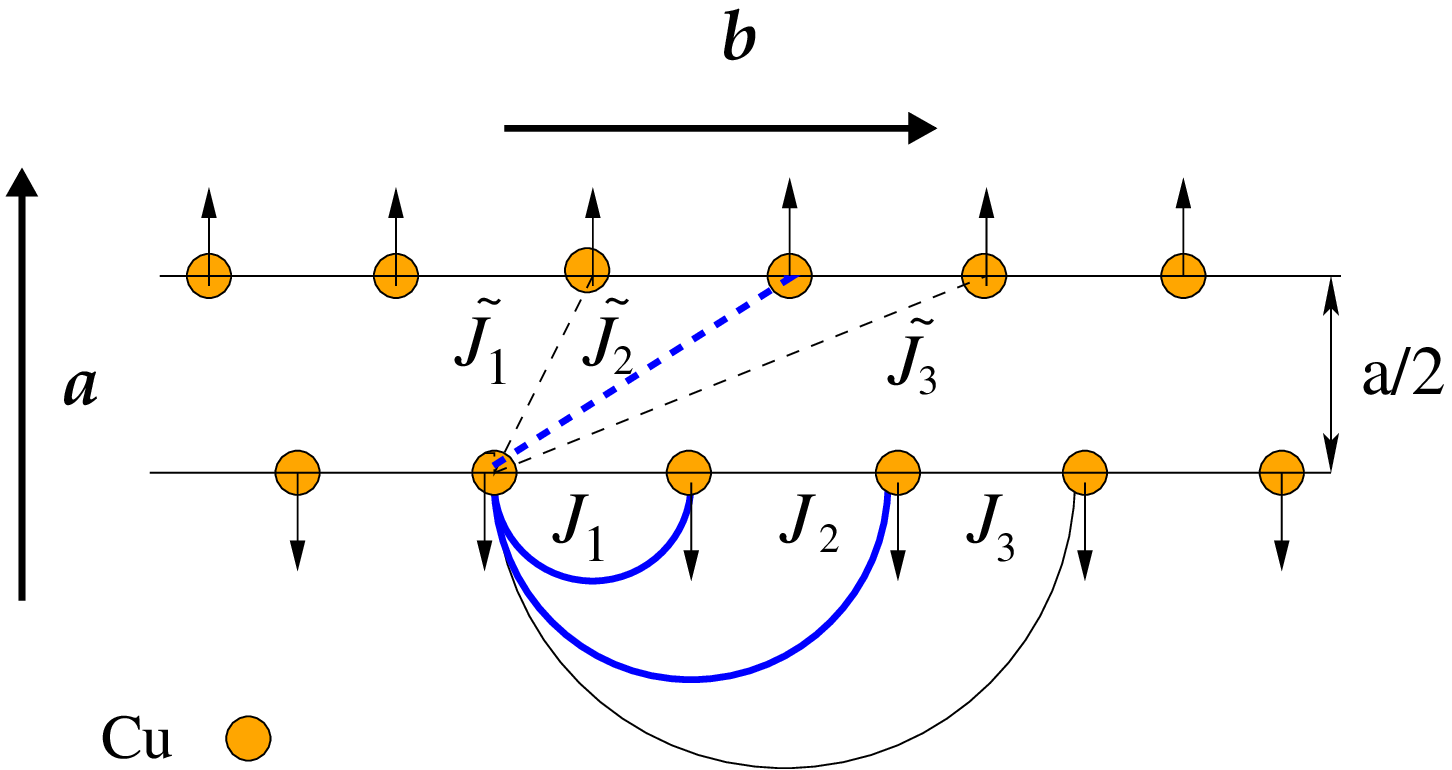}
\end{minipage}
\end{center}
\caption{(Color online) Left: The 
crystallographic structure of \Litwo\ comprises two AFM coupled
CuO$_2$ spin-chains per unit cell running  along the $b$-axis 
(orange {\large \color{orange} $\bullet$} -- Cu$^{2+}$, 
 red    {\large \color{red}    $\bullet$} -- O$^{2-}$, 
bright blue {\large \color{brblue} $\bullet$}
 -- Li$^{+}$). 
The unit cell is indicated by the outer black cuboid. 
Right: 
the main
 intra- and inter-chain exchange 
paths, $J_1$, $J_2$, and $\tilde{J}_2$ marked by
blue arcs and dashed lines, respectively. 
Notice the frustration introduced by an AFM inter-chain coupling
for  any non-FM in-chain ordering. 
} 
\label{fig::Struct}
\end{figure}
a critical evaluation of theoretical studies, especially of 
electronic/magnetic structure calculations
\cite{Mizuno1998,Weht1998,deGraaf2002,DrechslerJMMM2007,Xiang2007}, 
is still missing.  
\begin{figure*}[t]
\begin{minipage}{0.99\textwidth}
\begin{minipage}{0.49\textwidth}
\begin{flushleft}
\includegraphics[width=\textwidth]{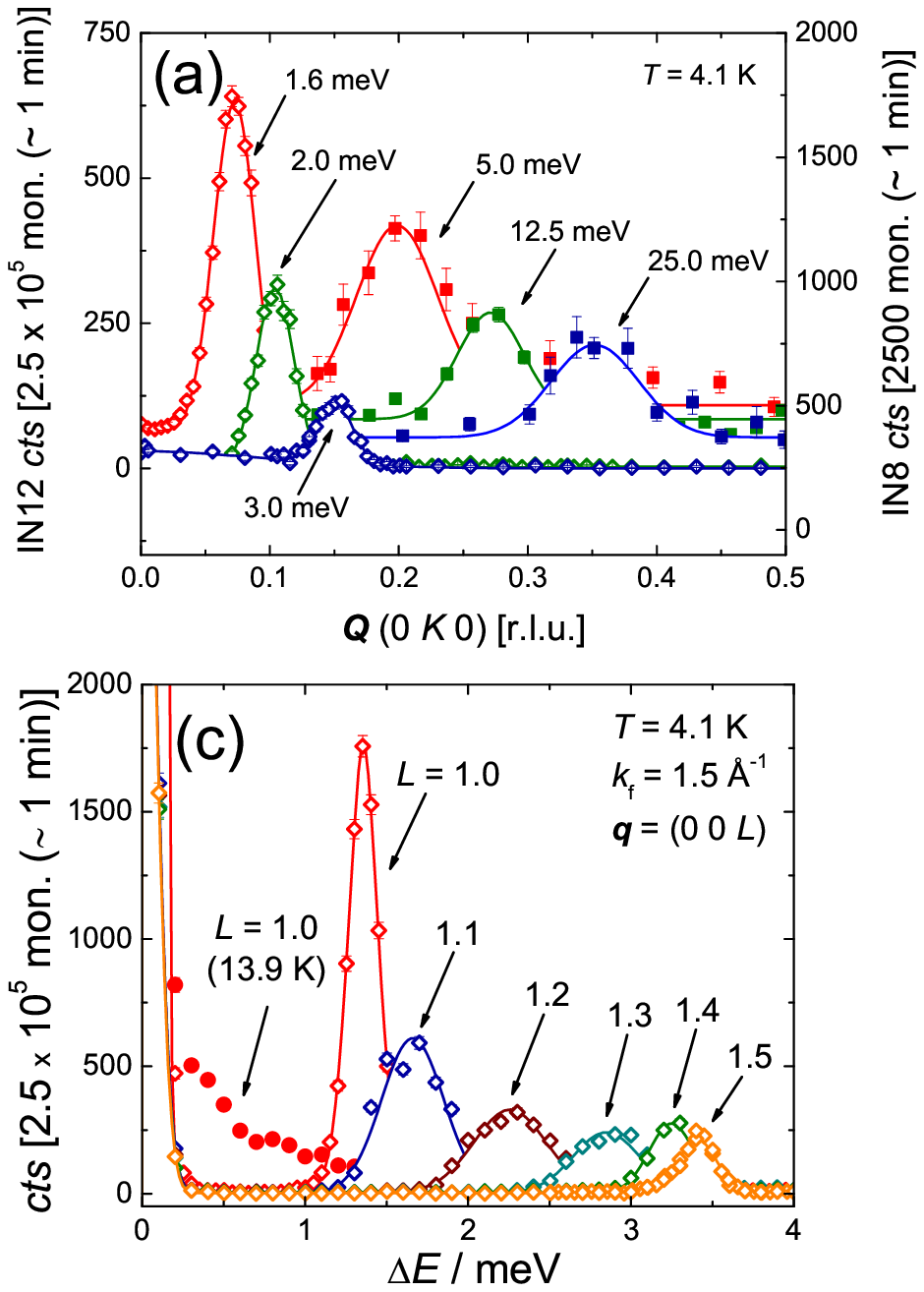}
\end{flushleft}
\end{minipage}
\hspace{0.75cm}
\begin{minipage}{0.40\textwidth}
\begin{flushright}
\includegraphics[width=\textwidth]{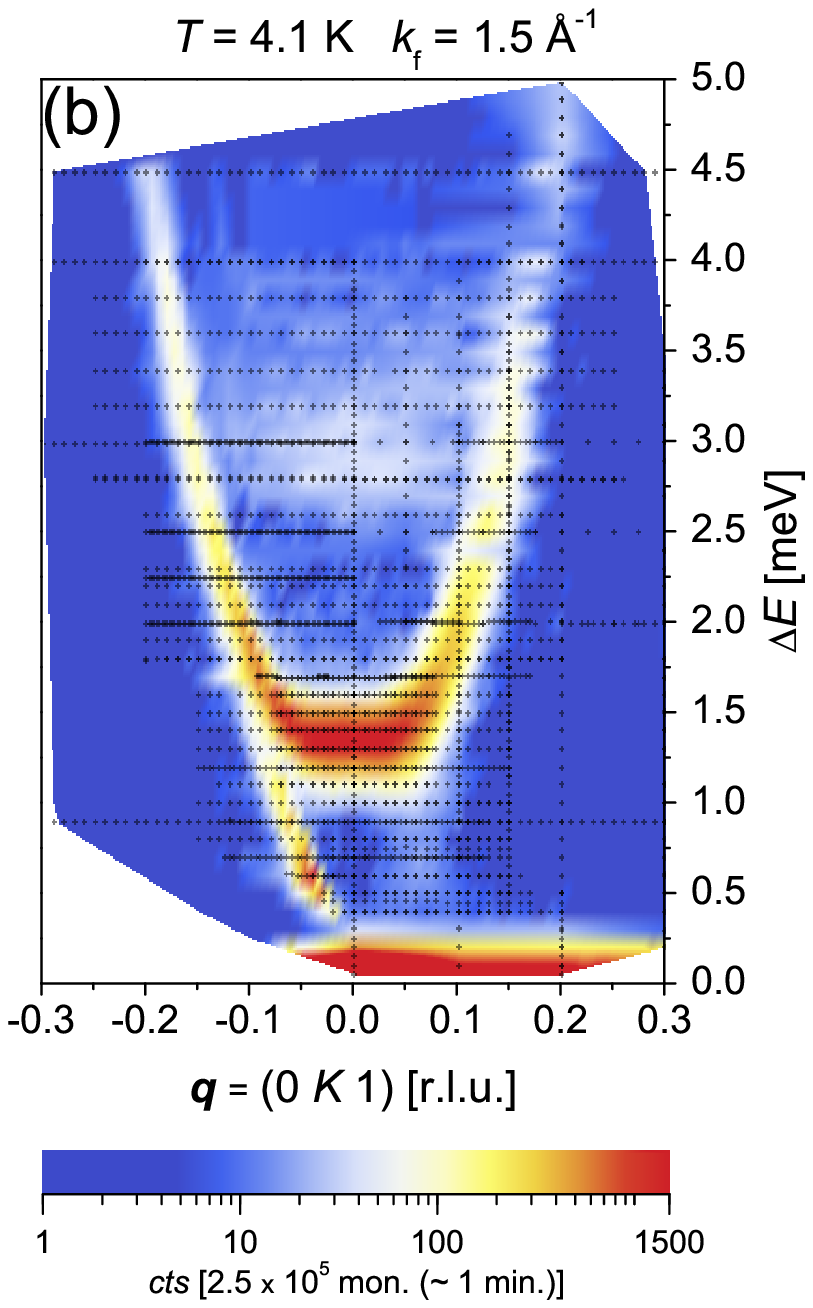}
\end{flushright}
\end{minipage}
\end{minipage}
\caption{(Color online) (a) Constant energy scans for momentum transfer 
along the chains with
$\mb{q}=(0$ $K$ $1)$ at IN12 ($\Delta E\le 3\meV$; open
symbols) and $\mb{q}=(0$ $1+K$ $0)$ for
IN8 data ($\Delta E \ge 5 \meV$; filled symbols). (b) Intensity map of 
the low energy spectrum, interpolated from the IN12 data points
({\small $\bullet$}). The sizeable intensity below the gap energy, i.e.\ at energy transfer
$\lesssim 1$\,meV and $q_b$  around -0.05 is a spurious
\textsc{Bragg} tail. 
(c) The dispersion perpendicular 
to the chains was measured by constant
$Q$-scans; lines are \textsc{Gauss}ian fits. Filled symbols: 
the measured intensity at the zone center for $T > \TN$.} \label{fig::INS}
\end{figure*}
In particular, this concerns
a precisely enough knowledge of the 
main exchange interactions.
The knowledge of realistic values is 
helpful also for the understanding of
related 
"frustrated ferromagnets" \cite{Dmitriev2008,Plekhanov2009} such as 
Ca$_2$Y$_2$Cu$_5$O$_{10}$ \cite{Matsuda2005}, 
La$_6$Ca$_8$Cu$_{24}$O$_{41}$ \cite{Matsuda1996} with 
FM in-chain ordering and LiVCuO$_4$, LiCu$_2$O$_2$ with
helimagnetism and multiferroicity, all being of
considerable current interest.

A previous inelastic neutron scattering (INS) study 
aimed to determine the
exchange integrals was not conclusive
\cite{Boehm1998}. It revealed an
anomalous low-lying branch of  hardly dispersive and overdamped  spin excitations 
in chain
direction. Its linear spin-wave (LSW)  analysis results in unrealistically small in-chain
exchange integrals. The missing, but expected,
dispersive 
quasi-1D spin chain excitation
remained as a challenging puzzle for the
community \cite{Weht1998,Mizuno1998,Mizuno1999}. 
In  Sec.\ 3 we
present new INS
data which unambiguously show the presence
of a strongly dispersive in-chain spin mode.
The main exchange integrals are derived 
applying the LSW-theory \cite{Oguchi1960} and  
in Sec.\ 4 
we compare our results
with those of related
chain cuprates as well as with  
predictions of band structure and cluster calculations.
A criticism of
improper \textsc{Curie-Weiss} analysis of 
spin susceptibility data 
is provided
and consequences for the direct FM Cu-O exchange parameter $K_{pd}$
entering extended \Hubbard{} 
models are discussed.

\section{{\bf 2. EXPERIMENTAL\/}}

A single crystal of $^7$\Litwo{} was grown for INS 
experiments by the travelling
solvent floating zone technique under high pressure \cite{Behr2008}. 
In order to avoid vaporization of 
Li$_2$O during growth a 4:1 Ar:O$_2$ atmosphere at $50\ab{\,\ab{bar}}$ was
chosen. A fast growth rate of $10\,\ab{mm/h}$ inhibits growth of impurity phases. Isotope enriched
$^7$Li was employed to avoid
the significant neutron absorption coefficient of $^6$Li. The sample
was characterized by X-ray powder diffraction, polarized light microscopy, magnetization and
specific heat measurements. By X-ray powder diffraction no impurity phase was found. The
macroscopic magnetization and specific heat data of the sample agree with literature data, i.e.
AFM order is found below $\TN = 9.2\K$ and a weak FM
component evolves below $T_2
\approx 3\K$ \cite{Ortega1998,Staub2000,Chung2003}.

INS experiments were performed with thermal and cold neutrons at the
three-axis-spectrometers IN8 and IN12 at the ILL, Grenoble, France. Four single crystals with
a total mass of 3.8\,g were mounted together
in the $(0$ $K$ $L)$ scattering plane with a resulting sample
mosaicity of $3^\circ$. For both instruments, focusing PG(002) monochromator and analyzer have
been utilized. The measurements at IN8 were taken with fixed final momentum
$k_\ab{f}=2.662\Angstrom^{-1}$ with PG-filter on $k_\ab{f}$. IN12 was configured with
$k_\ab{f}=1.5\Angstrom^{-1}$ and Be-filter on $k_\ab{f}$. Most scans have been done in the
CC-AFM-FM phase
at $T=4.1\K$ well above another not yet well understood magnetic phase 
below \tf\ . 
Anyhow, the observed changes of the INS spectra in this low-$T$ phase
(not shown here)
are weak.

\section{{\bf 3. RESULTS AND DATA ANALYSIS}}

The results of our INS studies are summarized in \figref{fig::INS}. Representative spectra of
constant energy scans as taken at IN12 and IN8 for moment transfer along the chains ($b^*$) are
displayed in \figref{fig::INS}(a). The main result is the observation of a highly dispersive
excitation which is strong at the magnetic zone center and significantly weakens at higher
energies. With the chosen experimental setup the magnetic branch could be traced up to energy
transfers of $25\meV$. The measured data points along $(0$ $K$ $1)$ taken at IN12 with cold
neutrons are summarized in the color map \figref{fig::INS}(b). Note, that the reflections are
periodic with the magnetic unit cell and their strongly  reduced intensity
above $\TN$ observed up to energy 
transfers of $15\meV$  does confirm their magnetic nature
(see Fig.\ 2 (c)).
At the magnetic zone center  $(0$ $0$
$1)$, a gap of $\Delta = 1.36\meV$ is observed. 
The excitations for momentum transfer along $(0$ $0$ $1+L)$ are only  weakly dispersive
in agreement with the results of Ref.\  
\cite{Boehm1998}. Respective
constant $\mb{q}$-scans are shown in \figref{fig::INS}(c). Note, that the mosaicity of the sample
broadens the excitations along $L$ which is less pronounced for moment transfer along $K$ due to
the longer $b^*$-axis.

As shown  in  
\figref{fig::INS}(b) we observe further inelastic features for moment
transfer along the chain. In addition, the data in \figref{fig::INS}(b) also exhibit weak and
presumably incommensurate (IC) magnetic scattering below the magnon gap energy. The origin of these
low-energy excitations is not yet clear and will be addressed in future studies. Furthermore,
there is a continuous feature appearing at double the energy of the anisotropy gap which
is possibly attributed to two-magnon scattering. We also mention that we have observed low-lying
and strongly broadened excitations along $b^*$ similarly as 
in Ref.\ \cite{Boehm1998}.
However, the intensity of these excitations was roughly two orders of magnitude weaker than
that reported ibidem.

According to ESR measurements \cite{Ohta1993} the exchange interactions 
show an uniaxial anisotropy with the easy-axis directed along 
the crystallographic $a$-axis. We describe the corresponding Cu momenta 
by the spin-\Hamilton{}ian
\begin{equation}
\hat{H}=\frac{1}{2}\sum_{\mathbf{m},\mathbf{r}}
\left[J_{\mathbf{r}}^{z}\hat{S}_{\mathbf{m}}^{z}\hat{S}_{\mathbf{m}
+\mathbf{r}}^{z}+J_{\mathbf{r}}^{xy}
\hat{S}_{\mathbf{m}}^{+}\hat{S}_{\mathbf{m}+\mathbf{r}}^{-} \right]\label{eq:H}
\end{equation}
where $\mathbf{m}$ enumerates the sites in the magnetic (Cu) lattice,
the vector  $\mathbf{r}$ 
connects sites with an exchange coupling
$J_{\bf r}$ \cite{remarkboehm}. The
$z$-axis is taken along the easy-axis, i.e.\ the $a$-axis. 
Within the LSW-theory \cite{Oguchi1960}, the 
dispersion-law 
reads
\begin{eqnarray}
\omega_{\mathbf{q}} & = &
\sqrt{\left(J_{\mathbf{q}}^{xy}-J_{\mathbf{0}}^{xy}+\tilde{J}_{\mathbf{0}}^{xy}
-D\right)^{2}-\left(\tilde{J}_{\mathbf{q}}^{xy}\right)^{2}},\label{eq:wq}\end{eqnarray}
where $J_{\mathbf{q}}\equiv(1/2)\sum_{\mathbf{r}}J_{\mathbf{r}}\exp\left(\imath\mathbf{qr}\right)$
is the Fourier transform of the in-chain exchange integrals,
and analogously for the inter-chain integrals $\tilde{J}_{\mathbf{q}}$.
The exchange anisotropy  
$D\equiv J_{0}^{z}-J_{0}^{xy}-\tilde{J}_{0}^{z}+\tilde{J}^{xy}_0$ causes the abovementioned spin
gap $\Delta =\omega_{\mathbf{0}}$ in our case 
(see Fig.\ 3).  Their relation reads
\begin{equation}
\Delta=\sqrt{D\left(D-2\tilde{J}_{0}^{xy}\right)},\quad \mbox {or} \quad
D=\tilde{J}_{\mathbf{0}}^{xy}-\sqrt{\left(\tilde{J}_{\mathbf{0}}^{xy}\right)^{2}+\Delta^{2}} \ .
\label{eq:D}
\end{equation}
In the summations over $\mathbf{r}$ we retain only 
the leading terms (see Fig.\ 1). According to
LSDA+$U$ based magnetic structure calculations they are given by the 
following in-chain integrals: $J_{1},\: J_{2},\: J_{3}$, (corresponding to
$\mathbf{r}=\mathbf{b},\:2\mathbf{b},\:3\mathbf{b}$ respectively) and  inter-chain integrals:
$\tilde{J}_{111},\:\tilde{J}_{131}$, (corresponding to $\mathbf{r}_{111}
=\left(\mathbf{a}+\mathbf{b}+\mathbf{c}\right)/2$,
$\mathbf{r}_{131}=\left(\mathbf{a}+3\mathbf{b}+\mathbf{c}\right)/2$).
For $q_aa$ or $q_cc=\pi$ the inter-chain dispersion caused by $\tilde{J}_{\mathbf{q}}^{xy}$ 
vanishes and Eqn.\ (\ref{eq:wq}) simplifies:
\begin{equation}
\omega_{\mathbf{q}}  = J_{\mathbf{q}}^{xy}-J_{\mathbf{0}}^{xy}+\tilde{J}_{\mathbf{0}}^{xy}
-D=J_{\mathbf{q}}^{xy}-J_{\mathbf{0}}^{xy}+\Delta_1 ,
\end{equation}
i.e.\ the single-chain dispersion can be read off {\it directly} 
by subtracting the effective gap $\Delta_1$, only.
 
Our INS data are well fitted by \textsc{Gauss}ian distributions. The maxima of the main
branch of the spectrum were analyzed within LSW-theory
\eqnref{eq:wq}. 
The inspection of Fig.\ 3  
\begin{figure*}[t]
\onefigure[width=.83\textwidth]{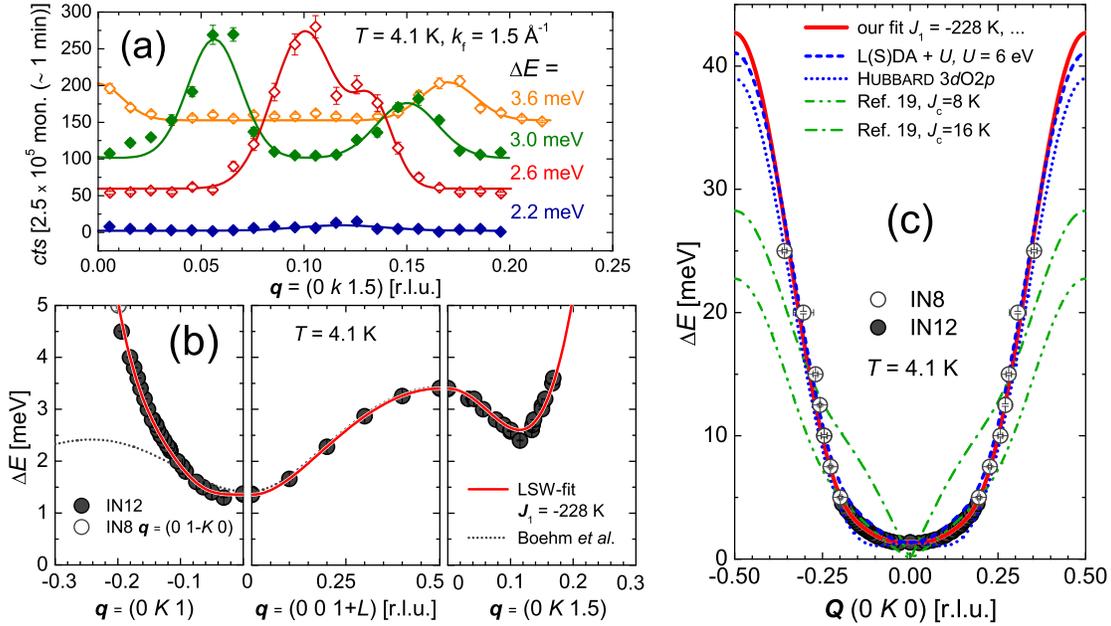}
\caption{(Color online) (a)-Constant energy scans for 
the data points shown in 
part (b) (right panel) near the minimum.
(b,c) - INS data points and LSW fits (red line)  of the magnon dispersion 
(b): along $\vec{q}
=$(0 $K$ 1), (0 0 $1+L$), (0 $K$ 1.5) for $E$-transfer below $5\meV$, compared 
with Ref.\ \cite{Boehm1998} (black dotted line);
(c) -
along
$Q=(0$~$K$~$0)$ up to high $E$-transfer. 
Solid red line: 
our fit (see Tab.\ 1).
For comparison the LSW dispersions predicted in Ref.\ \cite{Mizuno1999}
(green dashed-dotted and dashed double dotted lines),
by our L(S)DA+$U$ (blue dashed line) calculation and the
five-band \textsc{Hubbard} model (blue dotted line),
 both with the added
experimental spin gap
are shown, too. 
}
\label{fig::DispersionHigh}
\end{figure*}
reveals a strong in-chain dispersion  and a much weaker one in the 
perpendicular
$c$ and $a$ (not shown here) directions. The results of the fit are given in
\tabref{tab::ExchangeParameters}.
The total width $W=\omega_{\pi/b}-\Delta=
2\mid J_1+J_3+J_5+\cdots-0.5\left( \tilde{J}^{xy}_0-D \right) \mid -\Delta \approx 2\mid J_1\mid $, 
i.e.\ the in-chain dispersion yields  
a direct measure of the 
NN coupling since $W$ is {\it un}affected 
by $J_2, J_4, J_6 \cdots$.
Supposing a monotonous behaviour of $\omega_q$ up to
the zone boundary at
$Q=0.5$ (corresponding to $q=\pi/b$), from 
the measured part of the full $\omega_q$-curve one obtains
 already a 
rigorous lower bound
for $\mid J_1\mid \stackrel{>}{\sim}$ 150~K.  
Our analysis shows that the
in-chain 
NN
interaction $J_1$ is strongly FM, but 
frustrated by an AFM
NNN-coupling $J_2$
which affects the shape of $\omega_q$.
In comparison, the AFM inter-chain coupling is weak, clearly
demonstrating the magnetically quasi-1D
character of the compound.
For the in-chain coupling we find 
$\alpha = -J_2/J_1 = 0.33$, 
{\it unambigously above} the critical 
ratio $\alpha_{\ab{crit}}=1/4$ 
for an
isotropic 
\textsc{Heisenberg}-chain \cite{Bursill1995}.
Note that 
the dispersion near
the zone center behaves like $\omega(q)\propto q^4$
to be discussed below. 
We  confirm also 
theoretical predictions  
\cite{deGraaf2002,Xiang2007} 
(see also Tab.\ 1 for our results)
that the 
main inter-chain coupling is indeed the 
NNN
coupling along
$(\frac{a}{2},\frac{3b}{2},\frac{c}{2})$.
The NN inter-chain exchange has a negligible effect on the dispersion 
in the (0 $K$ $L$)-plane. Although further couplings 
 can not be 
 accessed from our fits,
the main $J$ values given here
do not change much, if the INS data are analyzed  in more complex models
with additional exchange
paths, especially $J_3$ and $\tilde{J}_1$. When taken into account, both remain small 
($\sim 4$~K and 1~K, respectively) in full accord with L(S)DA+$U$ (see Tab.\ 1).

\section{{\bf 4. DISCUSSION\/}}
The actual CC-AFM-FM ordering seemingly 
contradicts the IC "spiral" phase expected 
for a 
frustration ratio of $\alpha \approx 0.33$
in a 
1D-approach (in the sense of the wave vector $q_0\neq 0,\pi/b$ 
where the magnetic structure
factor $S(q)$ becomes maximal \cite{Bursill1995}).
Hence, 
the obtained relatively  small but frustrated AFM inter-chain
coupling may hinder the spiral formation. 
Thus, the 3D critical point compared with
$\alpha_c^{1D}$ is upshifted:
\begin{equation}
\alpha_c^{\rm \tiny 3D,iso}=\alpha_c^{1D}\left(1+\beta_1 +9\beta_2 +25\beta_3 + \cdots  \right) \ ,
\label{interchain}
\end{equation}
where $\beta_n= -\tilde{J}_n/J_1$. Eqn.\ (\ref{interchain}) has been derived in the 
isotropic (iso) case \cite{Drechsler2005}. 
Ignoring all other very weak inter-chain couplings $\tilde{J}_n$ 
we
arrive with our results  $\tilde{J}_2=9.04$~K and $J_1=-228$~K 
at
$\alpha^{\rm \tiny 3D, iso}_c=0.339$. Anisotropies 
(aniso),
as found here, further
stabilize the CC-AFM-FM state. 
From Eqns.\ (2,3) we estimate 
finally $\alpha_c^{\rm \tiny 3D,aniso}\approx 0.39$. 
The combined effect of 
AFM inter-chain coupling 
and easy-axis
anisotropy is also responsible for the anomalous 
$q^4$-dependence  
of the spin excitations mentioned above. 
In the limit $q_bb \rightarrow 0$ we expand Eqn.\ (2) and obtain
\begin{equation}
\omega(q)\approx\Delta +A_{\Gamma}(q_bb)^2+B_{\Gamma}(q_bb)^4, 
\end{equation}
\begin{table*}[]
\caption{The fitted exchange integrals (in K) as determined from the INS data using 
Eqns.\ (2,3) compared  
with microscopic theory (see text) and other 
recent theoretical results.
Values  in 
parentheses
 are estimates from less accurate fits.  }
\label{tab::ExchangeParameters}
\begin{center}
\begin{tabular}{r|c|c|c|c|c|c|c}
\hline\hline
& $J_1$        & $\alpha$           & $J_2=-\alpha\cdot J_1$ & $J_3$   &
$\tilde{J}_2$ &D         & $\tilde{J_1}$\\
INS / present work               & $-228\pm 5$ & $0.332\pm 0.005$   & $76\pm 2$             & (3.8)
& $9.04\pm 0.05$ & $-3.29\pm 0.2$ & (1) \\
\hline
3$d$O2$p$ / present work\cite{remarkmapping}:        & $-218$       & $0.30$             & $66$           
        & $-0.4$          & $-$              & $-$&$-$\\
3$d$O2$p$ \cite{Malek2008}:      & $-143$       & $0.23$             & $33$                   
& $-1$         &$-$ & $-$              & $-$\\
3$d$O2$p$ \cite{Mizuno1998}:     & $-103$       & $0.47$              
& $49$                   & $-2$         & $-$& $-$         & $-$\\
two-chain phenomenol. \cite{Mizuno1999}:     & $-100$       & $0.40$              
& $40$                   & $-$         & 16&  $-$        & 16\\
LSDA+$U$, $U=$~6 eV \cite{Drechsler2009}:   & $-216 \pm 2$            & 
0.31                  & $66\pm 2$     & $5\pm 2$
&$13\pm 2$&    $-$         & 0$\pm 2$\\
GGA +$U$, $U=$~6 eV\cite{Xiang2007}:            & $-171$           & 0.60                & 98
& $-$            & 18&     $-$        & 0.23\\
RFPLO, LAPW+SO\cite{Mertz2005}:            & $-$            & $-$                & $-$
& $-$            & $-$&     $-15.6 $       & $-$\\
\hline\hline
\end{tabular}
\end{center}
\end{table*}
$A_{\Gamma}$ and the quadratic dispersion vanish exactly
at 
\begin{equation}
\alpha^q_0=\frac{1}{4}\left(1+\frac{9\beta_2}{\delta}\right) = 0.33098, \quad 
\delta =1-\frac{D}{4\tilde{J}_2}.
\end{equation}
Accidentally \Litwo \ is very close to this point and its dispersion 
near the zone center is {\it quasi-quartic}.
But in the presence of a
spin gap $\Delta$ caused by the anisotropy $D$
this vanishing of the quadratic dispersion doesn't yet signal an instability 
of the CC-AFM-FM state. 

Near the Z-point $\left(0,0,\pi/c \right)$ the 
quadratic coefficient $A_Z$ is already essentially negative 
(see Fig.\ 3 (b, right panel)).
Since along the line $Z-R(0,\pi/b,\pi/c)$ the inter-chain dispersion
vanishes one can easily read off the 1D Fourier components
of the exchange interactions $J^{xy}_{\mathbf{q}}$ (see Eqn.\ (4)).
A similar rare situation  occurs  in the 2D 
frustrated CsCuCl$_4$ system 
in a high magnetic field above its saturation limit \cite{coldea2002}.
The clearly visible minima at  
$q_{b,0}b=\cos^{-1}\left(1/4\alpha \right)\approx\pm 0.72=\pm 0.11$~(r.l.u.)
correspond to the two equivalent 
propagation vectors of 
a low-lying spiral excitation \cite{Kuzian2007} 
with
a pitch angle of about 41.2$^{\circ}$
above a CC-AFM-FM  
ground state observed to the best of our knowledge for the first time.

Next, we briefly compare our results with those 
obtained so far by INS-studies  for 
Ca$_2$Y$_2$Cu$_5$O$_{10}$ (CYCO) \cite{Matsuda2005}
with a similar FM in-chain ordering and a
frustrating AFM inter-chain interaction. There the reported
 $J_1$ read
-80~K and -93~K for fits where $J_2=0$
and 
$J_2= 4.6$~K, respectively. However, 
such tiny
values of $J_2$
are
unlikely 
\cite{Mizuno1998,Mizuno1999,Malek2008}. For a 
standard Cu-O hybridization
a much larger
value is expected \cite{Kuzian2009} in accord with the
observed sizable part of the total magnetic moment 
(22 \% )
residing
at O \cite{Matsuda2002}.
Re-fitting
their INS data yields  
 $J_1$,$J_2$  
values of the same order as we found for \Litwo  \ (LCO) 
($J_1^{\tiny \rm CYCO}\sim J_1^{\tiny \rm LCO}$ and 
$J_2^{\tiny \rm CYCO}\sim 0.5J_{2}^{\tiny \rm LCO}$).
A detailed comparison of both systems
will be given elsewhere 
\cite{Kuzian2009}.
With respect to their large $J_1$ values  the question
may 
arise why they have been not recognised
so far in analyzing 
thermodynamic properties?
In this context a critical evaluation of the reported 
AFM "\textsc{Curie-Weiss}" (CW)
temperatures $\Theta_{\tiny \rm CW}^{\tiny \rm CYCO}\approx -15$~K \cite{Yamaguchi1999}
or small FM values: 5-10~K \cite{Kudo2005} and 
$\Theta_{\tiny \rm  CW}^{\tiny \rm LCO}\approx -40$ \cite{Sapina1990,Boehm1998} 
or -8~K \cite{Ebisu1998} 
is very instructive. All these data have been derived from the 
linear fits
\begin{figure}[!b]
\begin{center}
\includegraphics[width=0.29\textwidth,angle=-90]{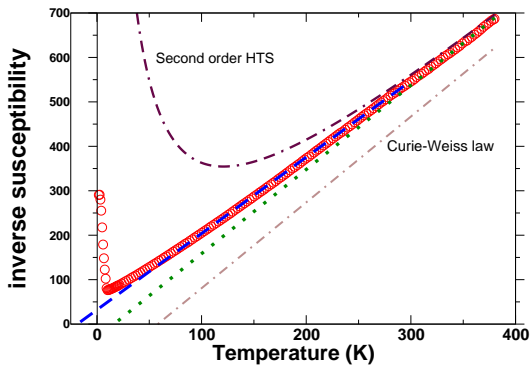}
\end{center}
\caption{(Color online) The 
inverse spin susceptibility as measured ($\circ$) 
vs high $T$-series 
expansion in 1$^{st}$ (correct CW-law)
and 2$^{nd}$ order. Dashed and dotted lines: "pseudo"-CW-laws.
} \label{fig::InvChi}
\end{figure}
of inverse susceptibility plots below 300--400~K
(of the type denoted as "pseudo"-CW-lines in Fig.\ 5). 
But here we estimate
$$\Theta_{\tiny \rm CW}\approx \frac{1}{2}\left[\mid J_1\mid-J_2-
\frac{z_{int-ch}}{2}\left(\tilde{J}_1+\tilde{J}_{2} \right) \right] > +54 \ \mbox{K},$$
where the inter-chain coordination number 
$z_{int-ch}=4,8$ for CYCO and LCO, respectively. Thus, we 
arrive at about $ 60$~K and 58$\pm 4$~K, respectively. 
The inspection of  \figref{fig::InvChi}
clearly shows that very high $T$
\cite{remarkCW}, {\it far above} any available data
would be required to extract $\Theta_{\tiny \rm CW}$ from  
$1/\chi(T)$-data. To satisfy the incorrect $\Theta_{\tiny \rm CW}$-values strongly 
underestimated $J_1$-values have been adopted in 
Refs.\ \cite{Yamaguchi1999,Kudo2005,Ebisu1998,Boehm1998,Mizuno1998}.  
Even the 2$^{nd}$ order of high $T$-series expansion (HTS) 
approaches the experimental curve only 
above 400~K. Hence, even more any attempt to
detect even a  \textsc{Curie}-law \cite{Mizuno1998} near 300~K 
must fail. Our $J_1$-values for LCO and CYCO from INS-data
provide support for the large value we found in Li$_2$ZrCuO$_4$ from
thermodynamic properties \cite{Drechsler2007PRL} but puzzle the tiny
value reported for LiVCuO$_4$ \cite{Enderle2005} 
with a similar Cu-O-Cu bond angle.

Finally, we turn to a microscopic analysis.
In Tab.\ 1 and Fig.\ 3 we compare our
INS derived exchange integrals with 
theoretical results. 
First, we list the in-chain couplings $J_n$ 
as obtained from the mapping of a
five-band extended Hubbard $pd$ model (on open chain Cu$_n$O$_{2n+2}$-clusters $n=5,6$) 
\cite{remarkmapping}
on a corresponding $J_1$-$J_2$-$J_3$-\Heisenberg{} model (see Fig.\ 1).
But here, to reproduce the main experimental
exchange integrals, a refinement has been
performed most importantly by considering
a larger direct FM  Cu-O exchange $K_{pd}=81$~meV compared with 50~meV 
adopted in Ref.\  
\cite{Mizuno1999}. We note that practically only $J_1$ is significantly affected by $K_{pd}$.
Thereby $\mid J_1 \mid \propto  K_{pd}$ holds approximately.
 Notice that the contribution of $K_{pd}$ is much more
important for the large negative (FM) value of $J_1$ than that of the intra-atomic
FM Hund's rule coupling on O. Since the 
available spectroscopic data at 300~K depend only weakly on $K_{pd}$ not
much is known on its magnitude. 
In the past $K_{pd}$ has been used mostly as a fitting
parameter for thermodynamic properties 
ranging from 50 to 110~meV for CuGeO$_3$ \cite{Mizuno1998,Braden1996}.
The INS data reported here provide a unique way to restrict its value phenomenologically
and opens a door for systematic studies of this very important 
interaction and
well-founded comparisons with other 
edge-shared CuO$_2$ chain compounds.

Secondly, in 
the LSDA+$U$ there is practically only one adjustable  parameter
$U_d-J_H$, where $U_d$ denotes the Coulomb onsite repulsion (between 6~and 10~eV) and $J_H$
denotes the intra-atomic exchange ($ \approx 1$~eV)
both on Cu-sites. Comparing the total energy
of various 
ordered magnetic states, a set of in-chain and inter-chain
integrals can be derived \cite{Xiang2007}. As a result one arrives again 
at very close numbers to our INS derived set
\cite{Drechsler2009}.
Noteworthy, both ED a well as the LSDA+$U$ provides a 
justification to neglect any long-range exchange
beyond the third NN. 
The latter also explains why 
there is only one
important inter-chain exchange integral $\tilde{J}_2$ ($\beta_2$). The excellent 
agreement between the
INS-data analyzed in the simple LSW-theory and the theoretical 
results/predictions suggests
that in the present case the effect of quantum fluctuations as well as of 
spin-phonon interaction seems to be
rather weak. The former point is also supported by the relatively large 
value of the magnetic
moment $m\approx 0.96\mu_{\tiny \rm B}$ \cite{Sapina1990,Chung2003}
in the ordered state below $T_N$ and a consequence of the fact 
that the FM state
is an eigenstate of the 1D spin-model in contrast to 
the \textsc{N\'eel} state. 
Concerning the value of the
anisotropy, there is no
good agreement between 
contemporary
DFT calculations \cite{Mertz2005}
and much smaller values obtained in
various experiments \cite{Ohta1993,Boehm1998}
including  our data (see Tab.\ 1).

\section{\bf 5.\ SUMMARY}
The main results
of our INS study are (i)
 the relatively large dispersion
of spin excitations in the 
CuO$_2$ chains  of \Litwo \ 
due to 
the
large value of the FM NN 
in-chain coupling $J_1$
and (ii) 
the observation of a 
low-energy
spiral excitation over 
a commensurate collinear \textsc{N\'eel} ground state in 
the vicinity of the 3D critical point
above the corresponding 1D point.
The obtained 
main exchange
integrals can be approximately reproduced adopting an enhanced
value for the direct FM exchange $K_{pd}$ 
between Cu 3$d$ and O 2$p$ states within an extended five-band
\Hubbard{}-model.
Further support for the empirical exchange 
integrals comes from L(S)DA+$U$ calculations, if a moderate value of 
$U$ somewhat smaller than the $U_d$ in 
exact diagonalization for
the extended 
\Hubbard{}-model 
is employed.
The achieved detailed knowledge 
of the main
exchange couplings 
derived from the INS-data
provides a good starting point for an improved
general theoretical description of other 
CuO$_2$-chain systems 
and to adress 
a
microscopic theory of
their
exchange 
anisotropy. 

\acknowledgments
\noindent
We thank the
DFG 
[grants KL1824/2 (BB, RK \& WEAL), 
DR269/3-1 (S-LD \& JM), \& 
the
E.-Noether-progr.\ (HR)],
the progr.\
PICS
[contr.\
CNRS 4767, NASU 243 (ROK)], and ASCR(AVOZ10100520) (JM)
for financial support 
as well as
M.\ Boehm, M.\ Matsuda, A.\ Boris, H.\ Eschrig,
V.Ya.\ Krivnov, D.\ Dmitriev,  
S.\ Nishimoto, E.\ Plekhanov
and J.\ Richter for valuable discussions.






\bibliography{99}

\end{document}